\documentclass[12pt]{iopart}
\usepackage{graphicx}
\usepackage{cite}

\usepackage{bm}
\usepackage{mathrsfs}
\usepackage{amssymb}
\usepackage{stmaryrd}

\newcommand{\eq}[1]{(\ref{#1})}
\newcommand{\fig}[1]{figure \ref{#1}}
\newcommand{\be}[1]{\begin{equation}\label{#1}}
\newcommand{\ee}{\end{equation}}
\renewcommand{\vec}{\mathbf}
\def\nix{}

\usepackage{graphicx}

\usepackage{mathrsfs}

\begin{document}

\title[The Coulomb four-body problem in a
classical framework]{The Coulomb four-body problem in a
classical framework: Triple photoionization of lithium}

\author{Agapi Emmanouilidou$^{1}$ and Jan
M.\ Rost$^{2}$} \address{$^{1}$School of Physics, Georgia Institute of
Technology, Atlanta, Georgia, 30332-0430} \address{$^{2}$Max Planck
Institute for the Physics of Complex Systems, D-01187 Dresden,
Germany} \date{\today}

\begin{abstract}
 Formulating a quasiclassical approach we determine the cross section
 for the complete four-body break-up of the lithium ground state
 following single photon absorption from threshold up to 220 eV excess
 energy.  In addition, we develop a new classification scheme for
 three-electron ionizing trajectories in terms of electron-electron
 collisions, thereby identifying two main ionization paths which the
 three electrons in the ground state of lithium follow to escape to
 the continuum.  The dominant escape paths manifest themselves in a
 characteristic ``T-shape'' break-up pattern of the three electrons
 which implies observable structures in the electronic angular
 correlation probability.  This break-up pattern prevails for excess
 energies so low that the Wannier threshold law $\sigma\propto
 E^{\alpha}$ describes already the triple ionization cross section,
 whose predicted value $\alpha=2.16$ we can confirm quantitatively.
\end{abstract}         

\maketitle

\section{Introduction}

 The broad interest in multi-electron photoionization processes is due
 to the fundamental role they play for understanding electron
 correlation induced by the long-range Coulomb forces.
 Identifying the different paths the electrons follow to escape to the
 continuum is essential in uncovering a variety of fundamental
 phenomena, from pattern in highly differential cross sections over
 interference phenomena to the energy dependence of the total
 ionization cross section.  The theoretical treatment of multiple
 ionization processes is highly complex with no analytic solution.  In
 the energy domain, the difficulty is that one has to account for the
 correlated motion of the electrons in the asymptotic form of the
 final continuum state.  In the time domain, this difficulty can be
 avoided at the expense of propagating the fully coupled few-body
 Coulomb problem in time.  
 
 For three electrons and sufficiently high photon energies this has
 been achieved in \cite{Pindzola04} and for even higher energies one
 may use approximate schemes \cite{Pattard}.  Experimentally, the
 first data on triple ionization appeared only recently \cite{Wehal98,
 Wehal00}, compared to data on photo double ionization which date back
 to the late sixties of the last century \cite{Ca67}.

In the current work, we present a theoretical study of the total
triple ionization cross section of lithium over a wide range in energy, from
threshold up to 220 eV excess energy, and predict characteristic
features in the electronic angular  correlation probability.
  While the total triple ionization cross section has already
been measured \cite{Wehal98} and compares favorably with our results,
the electronic angular correlation probability is not known experimentally.  However,
it should be measurable with state of the art experimental techniques.

Given the obstacles in the theoretical description stated above we are
only able to achieve these results by formulating the four-body
break-up process quasiclassically.  This implies classical
propagation of the Coulomb four-body problem using the classical
trajectory Monte Carlo (CTMC) phase space method.  CTMC has often been 
used to describe break-up processes induced by particle impact
\cite{CTMC1, CTMC2, CTMC3, CTMC4} with implementations differing
usually in the way the phase space distribution  of the
initial state is constructed.  We use a Wigner transform of the
initial quantum wave function for the initial state, and this is why
we call our approach ``quasi''-classical.  Naturally, the
electron-electron interaction is treated to all orders in the
propagation, and any difficulties with electron correlation in the
final state are absent, since the method is explicitly time-dependent.
%

While the classical results follow from a large numerical effort, they
still allow for a detailed analysis of the trajectories in terms of
their physical properties.  We will demonstrate that the triply
photoionizing trajectories can be organized in groups according to the
respective sequence of electron-electron collisions.  From the
emerging scheme we identify two main paths that lead to triple
ionization from the Li ground state.  The group to which an ionizing
trajectory belongs is identified in an automated process which
warrants a transfer of the classification scheme to more than
three-electron atoms without technical difficulties.  Physically, the
nature of the collision scheme also promises a generalization to more
electrons.

 Furthermore, being quasiclassical, our approach naturally addresses
 the energy regime close to threshold.  This energy regime has been
 traditionally of particular interest since the slow electron escape
 allows for large interaction times resulting in pronounced
 interactions among the escaping electrons.  We could confirm
 the Wannier threshold law $\sigma\propto E^{\alpha}$ for the
 four-body break-up of Lithium by single photon absorption with
 $\alpha=2.16$  in the energy range of 0.1-2 eV as detailed in 
 \cite{emro06}.
 
 The paper is organized as follows:  In section 2 we explain the 
 theoretical approach, in section 3 we present the most important 
 results, the triple ionization cross section and the electronic angular 
 correlation probability. Section 4 introduces the classification scheme of 
 ionizing trajectories which can explain the electronic angular 
 correlation probability in terms of the dominant ``T--shaped'' pattern of 
 the three escaping electrons. Section 5 is dedicated to a final 
 discussion and a summary.

\section{Quasiclassical theory of photoionization} We formulate the triple
photoionization process from the Li ground state ($1s^{2}2s$) as a two
step process \cite{Samson,Pattard,scch+02}.  First, one electron 
absorbs the
photon (photo-electron).  Then, due to the electronic correlations,
redistribution of the energy takes place resulting in three electrons
escaping to the continuum.  We express the above two step process as
  \begin{equation}
\label{prob}
\sigma^{3+} = \sigma_{\rm abs} P^{3+},\
\end{equation}  
where $\sigma_\mathrm{abs}$ is the total absorption cross section and
$P^{3+}$ is the probability for triple ionization.  In what follows,
we evaluate $P^{3+}$ and use the experimental data of Wehlitz
\cite{total} for $\sigma_\mathrm{abs}$.  Equally well, we could use a
theoretically calculated $\sigma_\mathrm{abs}$ \cite{emro06a} which is
easy to obtain following the approach of \cite{ro95}.  To compute
$P^{3+}$, we first assume that the photo-electron is a $1s$-electron. 
It absorbs the photon
at the nucleus ($\vec r_{1}=0$), an approximation that becomes exact
in the limit of high photon energy \cite{Kabir} and is in analogy to
the successful description of photo double ionization in two-electron
atoms \cite{scch+02}.  The photon
could also be absorbed by the Li 2s-electron.  However, the cross
section for photon absorption from a $1s$ orbital is much larger than
from a $2s$ orbital as investigated in \cite{emsc+03} for 
photoionization of an excited He($1s2s)$ atom.  Hence, we can safely assume
that the photo-electron is a $1s$ electron which significantly reduces
the initial phase space to be sampled.  Also, by virtue of their
different character the electrons become practically distinguishable
and allow us to neglect antisymmetrization of the initial state.  We
denote the photo-electron by 1, the other $1s$ electron by 2 and the
$2s$ electron by 3.  Immediately after photon absorption, we model the
initial phase space distribution of the remaining two electrons, $1s$
and $2s$, by the Wigner transform of the corresponding initial
wavefunction $\psi({\bf r}_{1}=0,{\bf r}_{2},{\bf r}_{3})$, where
${\bf r}_{i}$ are the electron vectors starting at the nucleus.  We
approximate the initial wavefunction as a simple product of hydrogenic
orbitals $\phi^{\mathrm{Z}_{i}}_{i}(\vec r_{i})$ with effective
charges $Z_{i}$, to facilitate the Wigner transformation.  The $Z_{i}$
are chosen to reproduce the known ionization potentials $I_{i}$,
namely for the 2s electron $Z_{3}=1.259$ ($I_{3}=0.198\,$a.u.) and for
the 1s electron $Z_{2}=2.358$ ($I_{2}=2.780\,$a.u.).  (We use atomic
units throughout the paper if not stated otherwise.)  The excess
energy, $E$, is given by $E=E_{\omega}-I$ with $E_{\omega}$ the photon
energy and $I=7.478$ a.u.\ the Li triple ionization threshold energy.
For a given $E$, the Wigner distribution $W$ has an energy spread and
it is only its expectation value that is equal to $E$ \cite{GeRo02}.
Since near $E=0$ energy conservation is vital we enforce it by
restricting the Wigner functions for the individual electron orbitals
to their respective energy shell \cite{cutoff}.  Following these
considerations, the initial phase space distribution is given by
\begin{equation}
\label{eq:distribution}
\rho(\Gamma) = \mathscr{N}
\delta(\vec{r}_1)\delta(\varepsilon_{1}+I_{1}-\omega)\prod_{i=2,3}W_{\phi^{\mathrm{Z}_{i}}_{i}}
(\vec r_{i},\vec p_{i})\delta(\varepsilon_{i}+I_{i})
\end{equation}
with normalization constant $\mathscr{N}$.

 We  determine the triple ionization probability $P^{3+}$ formally through
\begin{equation}
P^{3+}=\lim_{t\rightarrow\infty}\int_{t_{\mathrm{abs}}}^{t}{\rm d}\Gamma_{\mathcal
{P}^{3+}}\,
\exp((t-t_\mathrm{abs})\mathscr{L}_{\mathrm{cl}})\rho(\Gamma),
\label{eq:intphas}
\end{equation}
where the classical Liouvillian $\mathscr{L}_{\mathrm{cl}}$ is 
defined by
the Poisson bracket \{H, \} with  the full Coulomb
four-body Hamiltonian H  \cite{Ptriple} and the propagation begins 
at the time  $t_\mathrm{abs}$ of photoabsorption.  The projector $\mathcal P^{3+}$ indicates that
we integrate only over those parts of phase space that lead to triple
ionization.  Equation (\ref{eq:intphas}) amounts to propagating electron
trajectories using the classical equations of motion (CTMC).
Regularized coordinates \cite{regularized} are used to avoid problems
with electron trajectories starting at the nucleus.  
We evaluate $P^{3+}$
by weighting each triply ionized trajectory by the initial phase space
distribution and adding the contributions  \cite{scch+02}.

\section{Experimentally accessible observables: Total triple
photoionization and angular correlation probability} 

Observables calculated classically or semiclassically tend to better
approximate the ``exact'' value if they represent a quantity averaged
over as many degrees of freedom as possible.  The reason for this is
 that classically forbidden mechanisms such as tunneling
and pronounced interference effects are less likely to play a
prominent role if the dynamics is averaged over many degrees of
freedom.  Secondly, for classical calculations based on Monte Carlo
methods, one ``counts'' events, similarly as in the experiment.  This
means, only (randomly sampled) trajectories, whose final phase space
values fall into a certain bin, contribute to the observable.  As in
the experiment,  statistics of the contribution is an important
factor and therefore, more integrated observables are easier to
determine compared to  highly differential ones.  Interestingly the
classical domain of validity is complementary to that of approximate quantum
calculations, e.g., with the so called 6C wavefunction \cite{maro+97}
where the fully differential cross section is calculated and more
integrated observables require numerically expensive
integrations.

For the present case, this implies that our focus is the triple
ionization cross section (which is the ``most integrated''
three-electron observable) with our statistics also allowing the
evaluation of single differential cross sections.  Here, we
present for reasons which will become clear in section 3 an unusual
but observable quantity, the electronic angular correlation
probability, which answers the question: How likely is it in triple
ionization to find two electrons under a certain angle $\theta$?

\subsection{Triple photoionization cross section}

\begin{figure}
\includegraphics[height=6cm,clip=true]{li-crossaa.eps}
\includegraphics[height=6cm,clip=true]{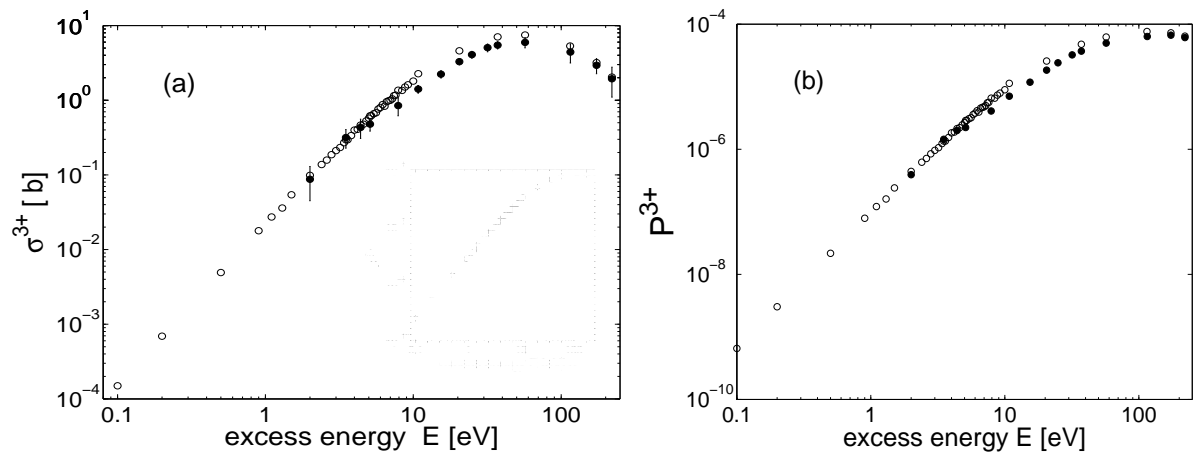}
\caption{\label{fig1:li-crossProp} (a) Triple photoionization cross section
obtained by multiplying the triple photoionization probability  from the present
calculation ($\circ$) with the total photo cross section from \cite{total} 
in comparison to the experiment \cite{Wehal98,Wehal00} (experimental
results ($\bullet$) shown with error bars). (b) shows triple photoionization
probability $P^{3+}$ as a function of excess energy: present
calculation ($\circ$); experiment ($\bullet$) \cite{Wehal98,Wehal00,total}.
}
\end{figure}

We have already described in section 2 how to determine the triple photo 
ionization cross section $\sigma^{3+}$ which is shown in 
figure \ref{fig1:li-crossProp}a. Note that no fitting parameters are 
used to obtain $\sigma^{3+}$. However, one may object that we use 
the total photo cross section extracted from another calculation or 
the experiment  for the curve in figure \ref{fig1:li-crossProp}a.
Hence we present in figure \ref{fig1:li-crossProp}b the triple 
ionization probability $P^{3+}$ which we calculate directly and which 
can be also obtained directly from experimental data using the 
relation
\be{ratios}
P^{3+}=R^{3+}/(1+R^{2+}+R^{3+})\,,
\ee
where $R^{3+}$ and $R^{2+}$  are the
experimental triple-to-single and double-to-single photoionization
ratios, respectively \cite{Wehal98,Wehal00,total}.
Figure \ref{fig1:li-crossProp}b also illustrates  the
numerical effort involved in computing $P^{3+}$.  For example, to
obtain $10^{3}$ triple photoionizing trajectories at $E=0.9$ eV with
$P^{3+}\sim 10^{-7}$ one has to evolve $10^{10}$ trajectories with the
CTMC method \cite{CTMC1}.

 Considering the approximations we had to make to handle the four-body
 problem, the agreement with the experimental data is remarkably good,
 starting near threshold where we can confirm the classically expected
 behavior of the cross section \cite{emro06} according to Wannier's
 theory \cite{Wan53,Klar}.  The agreement extends beyond the maximum
 of the cross section where our results also agree with the data
 points recently obtained in an ab-initio calculation
 \cite{Pindzola04}.  Hence, our classical approach with an approximate
 initial quantum wavefunction apparently captures the relevant
 correlations among the three electrons.  For very high excess
 energies (currently not considered) the triple photoionization can
 not be adequately described by our quasiclassical formulation.  However, one can describe the
 process using the Born approximation.

\subsection{Angular correlation probability}
\begin{figure}
\begin{center}
\includegraphics[scale=0.5,clip=true]{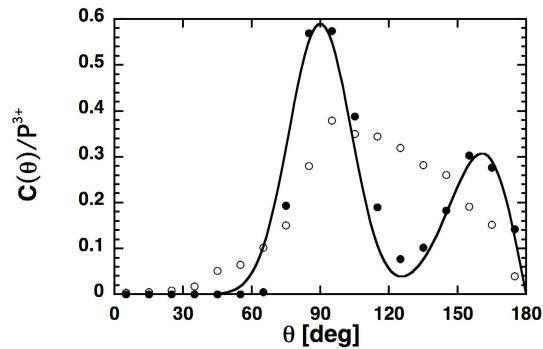}
\caption{\label{fig4:li-angles-all} The angular correlation 
probability 
$C({\theta})/P^{3+}(E)$ defined in \eq{theta}  for $E=0.9$ eV 
($\bullet)$
and  for $E=220.5$ eV ($\circ$), normalized to the respective 
triple ionization probability ($P^{3+}(E=0.9\,$eV$) = 7.89\times 10^{-8}$,
$P^{3+}(E=220.5\,$eV$) = 6.43\times 10^{-5}$). 
The number of events $N$ has been binned over $10^{\circ}$
at $\theta_{j}= 5^{\circ}+j 10^{\circ}$, $j = 0,1,\ldots,17$.  
For the solid line, see section \ref{sec:4.3}.}
\end{center}
\end{figure}

As already indicated, with only $10^{3}$ triple ionization events out
of $10^{10}$ trajectories, at most a single differential observable
can be determined.  Interesting with respect to dynamical correlation
among the three electrons and experimentally accessible is the angular
correlation probability
\be{theta}
C(\theta)
= \lim_{t\rightarrow\infty}\sum_{i>
j=1}^{3}\int_{t_{\mathrm{abs}}}^{t}{\rm d}\Gamma_{\mathcal {P}^{3+}}\,
\exp((t-t_\mathrm{abs})\mathscr{L}_{\mathrm{cl}})\rho(\Gamma)
\delta(\theta_{ij}(t)-\theta)
\ee 
with
\be{theta-2}
\theta_{ij}(t)=\arccos[\vec p_{i}(t)\vec p_{j}(t)/(p_{i}(t)p_{j}(t))]
\ee
which  depends only on
the \textit{relative} angle $\theta_{ij}$ between any pair of ionized
electrons in the three electron escape.  Formally, this is easily
achieved within a CTMC approach.  However, the number of events is
very small and we need to bin the observable over 10$^{\circ}$
respectively, as shown in figure \ref{fig4:li-angles-all}.  We see for
the higher excess energy ($E=220.5$ eV) a broad distribution with a
maximum near $90^{\circ}$.  This might have been expected from
impulsive binary collisions of the fast photo electron (peak at $\theta = 90^{\circ}$) where the
shift to slightly larger $\theta$ indicates the (small) influence of
the Coulomb repulsion.  At $E=0.9$ eV, the situation is drastically
different: A double-hump structure emerges with peaks at 
$90^{\circ}$ and $180^{\circ}$\footnote{Convolution with the
volume element $\sin\theta$ leads to an appearance of the
$180^{\circ}$--peak
at a smaller angle, see section \ref{sec:4.3}.}
This can be interpreted as ``T-shaped'' structure
which the three outgoing electrons form, where two of them leave along
a line towards opposite sides and the third one leaves perpendicularly
to the line.  The origin of this double-hump structure and its
interpretation as a T-shape configuration of the escaping electrons
will become clear after the analysis of the electron collision
sequences in the next section.

\section{Classification of triple ionization dynamics in terms of an
electron-electron collision scheme}

\subsection{The definition of electron collisions in a multi-electron 
system}

 The (dipole) coupling to the electromagnetic field
is a single-electron operator. Hence, one electron
absorbs the photon, which gets thereby annihilated.  Before this photo
electron has left the atom it must transfer part of its energy to the
other electrons to be ionized within a very short time.  How this
happens, directly or indirectly, is a question which is difficult to
ask quantum mechanically, but is a natural question in classical
mechanics, where the electrons undergo soft collisions mediated by
Coulomb forces.  In contrast to billiard balls they can indeed transfer
energy among themselves if the nucleus is within the reach of the
Coulomb potential to absorb the recoil momentum.  Moreover, in a
two-electron atom, it is most likely that a single collision among the
two electrons occurs which does not provide a lot of insight.
However, in three-electron atoms, the situation is far more complex
and a priori the nature of the collisions is not clear: Is triple
ionization mediated mostly by a single collision involving all three
electrons or does it happen sequentially with a sequence of
momentum transferring two-electron collisions?  If the latter is the case, is there a
pattern of preferred sequences?  Does a characteristic collision
pattern of classical trajectories lead to observable consequences?

\begin{figure}
\includegraphics[width=0.5\textwidth]{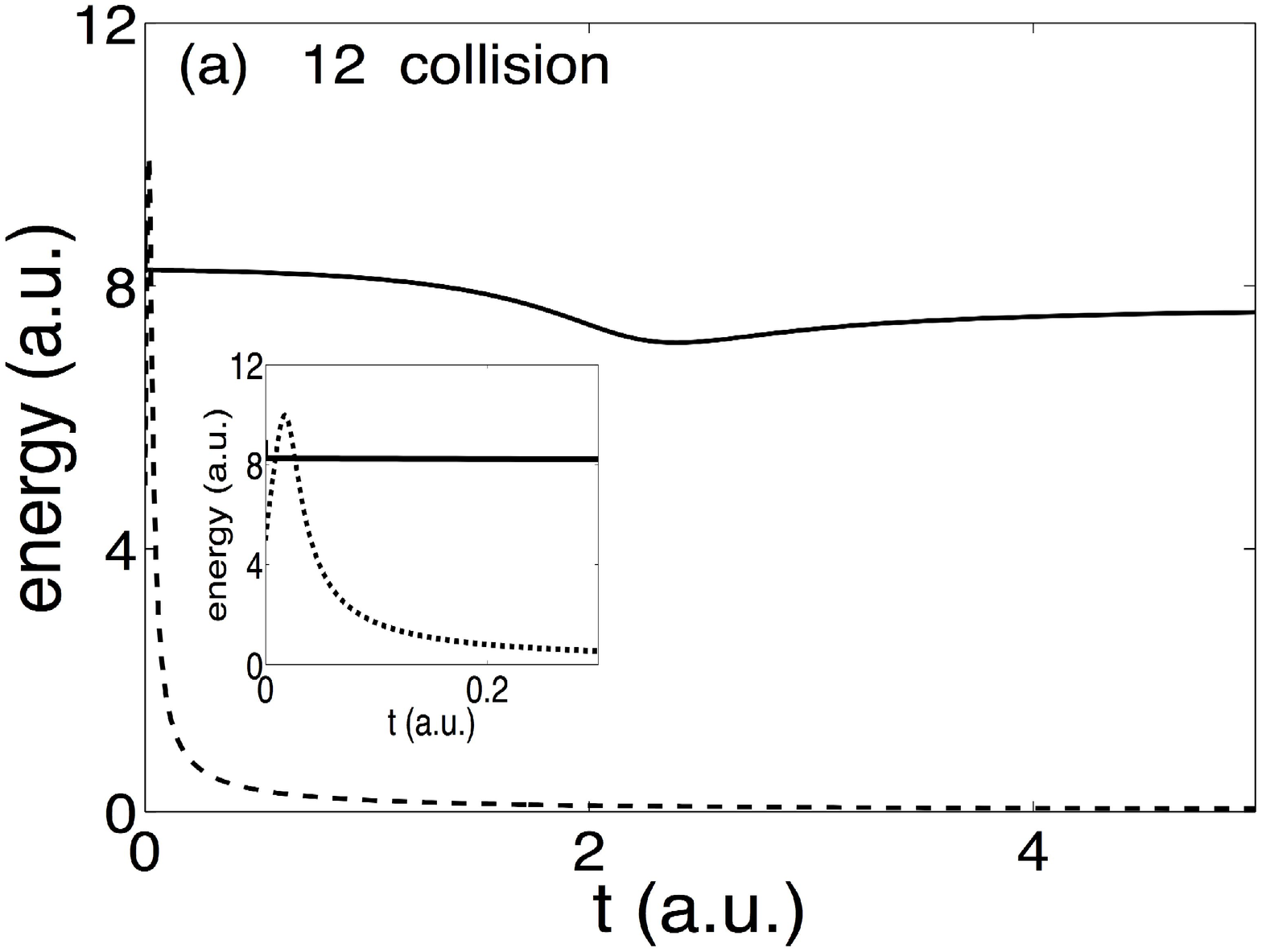}
\includegraphics[width=0.5\textwidth]{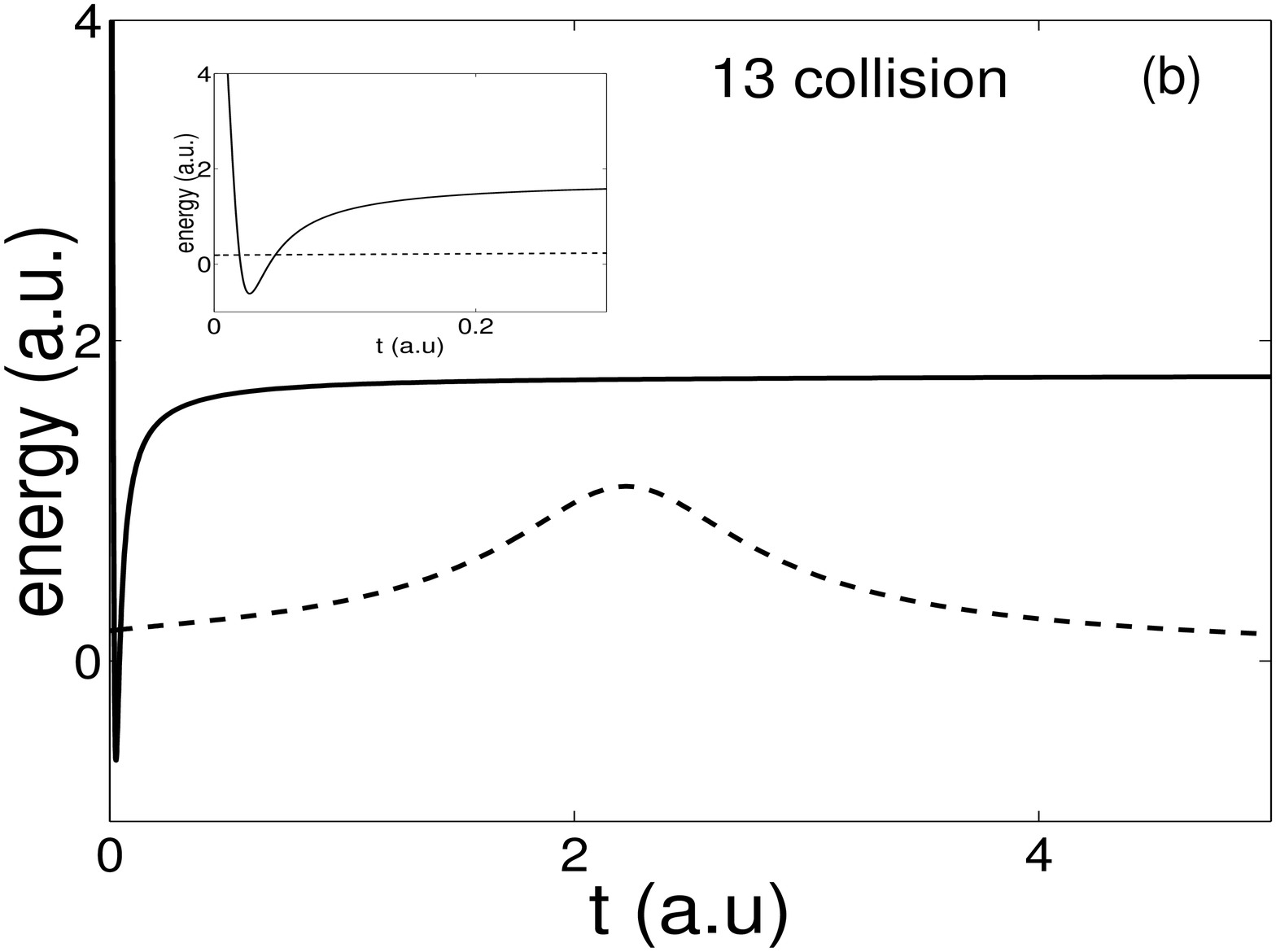}
\caption{\label{fig5:H12132205} Energies in time for three-body
subsystems for a triple ionizing trajectory 
with (1$\nix$2,1$\nix$3) electron-electron
collision sequence at $E=220.5$ eV. The maxima of V$(r_{12})$ and
V$(r_{13})$ occur at times $0.017$ a.u. (0.41 attoseconds) and $2.22$ 
a.u. (54 attoseconds), respectively.
In (a) the solid line is the energy of the H$_{12}$ subsystem, while
the dashed line is the interaction V$(r_{12})$.  In (b) the solid line
is the energy of the H$_{13}$ subsystem, while the dashed line is
V$(r_{13})$. The insets are for short times $t$.}
\end{figure}

Before we can answer these questions we first have to define what we 
call a momentum transferring electron-electron collision along a 
trajectory with time dependent electron positions $\vec 
r_{i}(t),\,i=1,2,3$.
The term
  responsible for momentum transfer between electrons $i$ and $j$ is
 their Coulomb repulsion $V(r_{ij})=r^{-1}_{ij}$, $\vec r_{ij}=\vec
  r_{i}-\vec r_{j}$.  
Hence, we identify a collision between electron $i$ and $j$  ($i\nix j$)
 through the momentum transfer 
 \be{mom}  \vec D_{ij}: =
 -\int_{t_{1}}^{t_{2}}\nabla V(r_{ij})\,dt 
 \ee 
 under the condition
 that V$(r_{ij}(t_{k}))$, $k = 1,2$ are local minima in time with
 $t_{2}>t_{1}$, while $r_{ij}=|\vec r_{i}-\vec r_{j}|$.  This automatically ensures that the integral of
 \eq{mom} includes the ``collision'' with a local maximum of
 V$(r_{ij}(t_{k}))$ at a time $t_{1}<t_{M}<t_{2}$.  During the time
 interval $t_{1}<t<t_{2}$, all four particles interact with each
 other. Hence, the   definition \eq{mom} is only meaningful if 
  the collision  redistributes energy dominantly
 within the subsystem given by the two-electron Li$^{+}$-Hamiltonian
 H$_{ij}$ of the nucleus and the electrons $i$ and $j$ involved in the actual
 collision.  This is indeed the case since the energy in the subsystem H$_{ij}$
 changes little over the collision, i.e.,
 $\int_{t_{1}}^{t_{2}}dH_{ij}/dt\,dt \ll E$, where E is the total
 energy of the Li-Hamiltonian.  We illustrate the latter statement in
 figure \ref{fig5:H12132205} using as an example a triple ionizing trajectory
 labeled by the sequence of (1$\nix$2,1$\nix$3) electron-electron
 collisions.  During the 1$\nix$2 collision (see figure 
 \ref{fig5:H12132205}a),
 V$(r_{12})$ undergoes a sharp change while H$_{12}$ changes smoothly.
 Thus, during the 1$\nix$2 collision, the V$(r_{12})$ potential
 energy is primarily redistributed in the Li$^{+}$ subsystem of
 electrons 1 and 2.  Similarly, during the 1$\nix$3 
 collision (see figure 
 \ref{fig5:H12132205}b), V$(r_{13})$ undergoes a sharp change while $H_{13}$
 changes smoothly.  Thus, the V$(r_{13})$ potential energy is
 primarily redistributed in the Li$^{+}$ subsystem of electrons 1 and
 3.  On the other hand H$_{13}$ and H$_{12}$ undergo a sharp change during
 the 1$\nix$2 and 1$\nix$3 collisions, respectively.  This
 should be the case, since during these times it is the energy of the
 H$_{12}$ and H$_{13}$ subsystem, respectively, that is conserved.
 Note that the higher the excess energy is, the more impulsive the
 electron-electron collisions are, since the collision time becomes
 shorter and shorter compared to the time the bound electrons need to
 orbit around the nucleus.

\subsection{Dominant collision sequences}

\begin{figure}
\centerline{
\includegraphics[scale=0.5,clip=true]{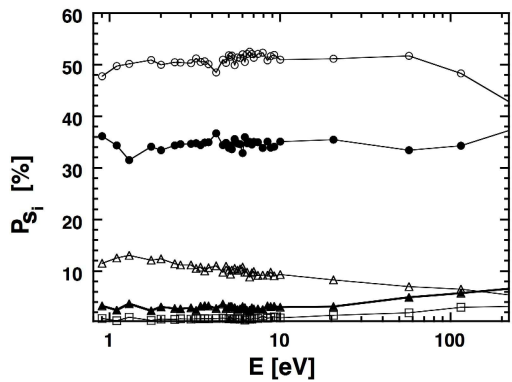}
}
 \caption{\label{fig3:li-mechanisms} Probability relative to all triple ionizing
 trajectories classified according to the sequence $s_{i}$ of
 electron-electron collision i$\nix$j, see text; $s_{1}$ =
 (1$\nix$2,1$\nix$3) ($\circ$), $s_{2}$ = (1$\nix$2,2$\nix$3)
 ($\bullet$), $s_{3}$ = $(13,12)$ ($\square$), $s_{4}$ = 
 (1$\nix$2,1$\nix$3,2$\nix$3) 
 ($\triangle$), $s_{5}$ = (2$\nix$3,1$\nix$2,1$\nix$3) ($\blacktriangle$).  }
\end{figure}

For the majority of triple ionizing trajectories, we register at least
two electron-electron collisions.  For automated identification of the
collisions sequences we need a sensitivity threshold to register only
the important collisions for the triple ionizing trajectories.  This
is done individually for each trajectory by forming the maximum $D =
\max_{i\ne j}\{|\vec D_{ij}|\}$ and normalizing each collision
according to $D_{ij}\equiv |\vec D_{ij}|/D$.  A collision is only
registered if $D_{ij}> \delta$.  The resulting classification
shown in figure \ref{fig3:li-mechanisms} does not sensitively depend
on the exact value of $\delta$ which we have chosen to be $\delta =
1/8$.

 According to our classification scheme the photo-electron transfers
 energy to the other two electrons through two main sequences of
 collisions in about 84\% of all triple ionization events.  In the first
 main pathway to triple ionization, the
 $s_{1}$=(1$\nix$2,1$\nix$3) sequence, the photo electron ($1$)
 knocks out, successively, electrons 2 and 3.  In the second main
 pathway, the $s_{2}$=(1$\nix$2,2$\nix$3) sequence, the
 photo-electron 1 first knocks out electron $2$.  Then, electron $2$ knocks out
 electron $3$.  It is easy to understand how the
 relative probability of these two processes is changing as a function
 of excess energy.  The process $s_{1}$=(1$\nix$2,1$\nix$3) has
 the highest weight for low excess energy where the photo-electron
 (after photon absorption) is still slow enough to easily transfer
 energy, first to electron 2 and then to electron 3.  For higher energy, the
 competing process $s_{2}$=(1$\nix$2,2$\nix$3) takes over because
 the photo-electron is still so fast after its first collision with
 electron 2 that the interaction with the more loosely bound 2s
 electron 3 is small.  Rather, it is more probable that electron 2
 transfers part of the energy it has gained after the initial
 1$\nix$2 collision to electron 3 through the 2$\nix$3 collision.
 
 Generally speaking, the  more collisions a sequence has, the less  is
 its weight  which can be seen in figure \ref{fig3:li-mechanisms} 
 comparing the weight of $s_{4}$ and $s_{5}$ with $s_{1}$ and 
 $s_{2}$. However, there are also exceptions, as illustrated by
 $s_{3}=(1\nix 3,1\nix 2)$. That this sequence does not carry large 
 weight is obvious since it is not very likely that the photo-electron  
 collides first with (2s) electron 3 and later with the more tightly 
 bound (1s) electron 2.


\subsection{The influence of the collision scheme on the 
angular correlation probability}
\label{sec:4.3}
\begin{figure}
\centerline{
\includegraphics[scale=.3,clip=true]{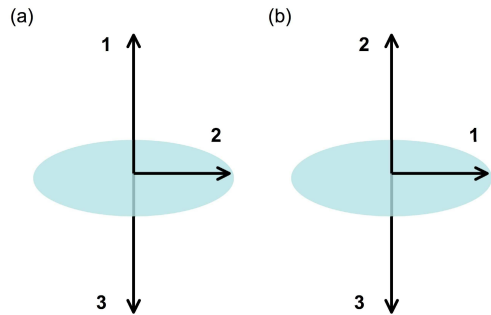}
}
\caption{\label{fig:tshape} Sketch of the T-shape structure when the
three electrons escape to the continuum through the $s_{1}$ process
(a) and through  the $s_{2}$ process (b).}  
\end{figure}

\begin{figure}\begin{center}
\includegraphics[scale=0.7,clip=true]{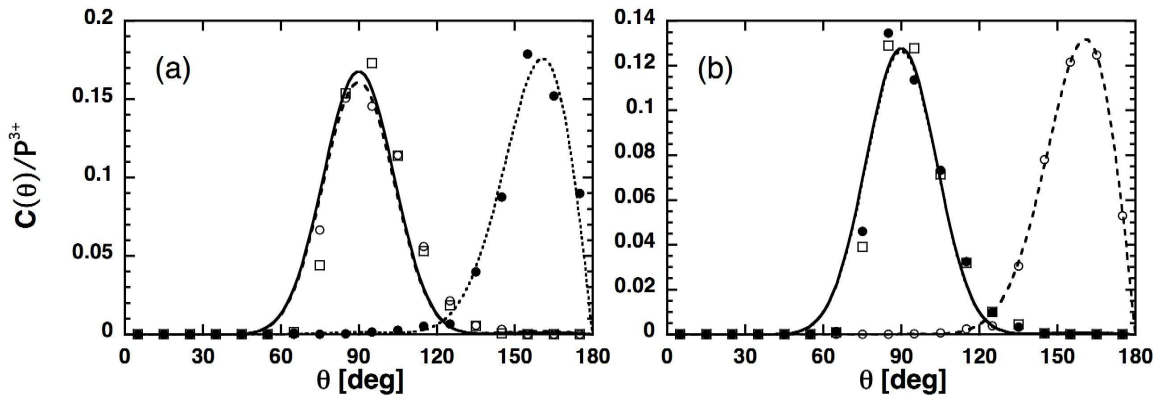}
\caption{\label{fig5:li-angles-resolved} Angular correlation 
probability for $E=0.9\,$eV as in \fig{fig4:li-angles-all}, but  
between two specific electrons
12 (open square/solid line), electrons 13 (filled circle/dotted line) and 
electrons 23 (open circle/dashed line). Part (a) is for trajectories from 
the $s_{1}$ sequence, part (b) for those from the $s_{2}$ sequence, 
for details, see text.
}
\end{center}
\end{figure}

One would expect that the collision scheme with its two dominant
pathways to the continuum, the sequences $s_{1}$ and $s_{2}$, has a
prominent influence on collisional observables.  Indeed, as we will
show, it explains the T-shaped configuration (see figure \ref{fig:tshape})
of the three outgoing electrons which manifests itself in the angular
correlation displayed in figure \ref{fig4:li-angles-all}. 
To this end, we have investigated the angular correlation $C(\theta)$
in more detail by asking which angle $\theta_{ij}$ the individual
electron pairs ij (i.e., 12, 23, and 13) form upon leaving the
nucleus.  This allows us to determine the characteristic $C(\theta)$ a
specific collision sequence $s_{i}$ produces. 
We
show the result in figure \ref{fig5:li-angles-resolved} for the two
dominant sequences $s_{1}$ and $s_{2}$.  One sees that the electron
pair which undergoes the \textit{last} collision in a sequence (13 for
$s_{1}$ and 23 for $s_{2}$) leaves towards opposite sides with an
angle of $\theta = 180^{\circ}$.  On the other hand, the pair which
collides first (12 for both sequences) forms an angle of $\theta =
90^{\circ}$, and so do the pairs which do not undergo an explicit
collision (23 for $s_{1}$ and 13 for $s_{2}$.  
The solid and dashed lines in figure
\ref{fig5:li-angles-resolved} are fits with the functions
\begin{eqnarray}
    \label{senkrecht}
    C^{ij}_{\perp}(\theta)& = &c_{\perp}^{ij}\sin^{\beta_{\perp}}\theta\sin\theta\\
    \label{parallel}
    C^{ij}_{\parallel}(\theta)& = 
    &c_{\parallel}^{ij}\sin^{\beta_{\parallel}}\theta/2\sin\theta\,,
    \end{eqnarray}
respectively, where the $c^{ij}$ and $\beta$ are fitting parameters,
while $\sin\theta$ comes from the line element of integration,
$\sin\theta d\theta$.  Taking the direction of the electron leaving 
in the plane perpendicular to the pair escaping back to back as a 
reference, it is easy to see that the average width in the 
distribution about $180^{\circ}$ should be twice the one about 
$90^{\circ}$, i.e., $\beta_{\parallel}\approx 
2\beta_{\perp}\equiv 2\beta$. This is indeed the case and all curves in
\fig{fig5:li-angles-resolved} are fitted well by $\beta = 16.8$.
The  fits to the individual curves in 
\fig{fig5:li-angles-resolved} together produce the curve
shown in \fig{fig4:li-angles-all} according to
\begin{equation}\label{fittheta}
C(\theta) = \sin\theta[c_{\perp}\sin^{\beta}(\theta) + 
c_{\parallel}\sin^{2\beta}(\theta/2)]
\end{equation}
with
\begin{eqnarray}
    \label{senkrecht1}
    c_{\perp}& = &c_{\perp}^{12}(s_{1})+c_{\perp}^{23}(s_{1})+
    c_{\perp}^{12}(s_{2})+c_{\perp}^{13}(s_{2})\\
    \label{parallel1}
    c_{\parallel}& = 
    &c_{\parallel}^{13}(s_{1})+c_{\parallel}^{23}(s_{2})\,.
    \end{eqnarray}
For the two dominant collision sequences $s_{1}$ and $s_{2}$
the T-shape geometry implies that the inter-electronic angle
$90^{\circ}$ is produced twice (by the first pair of electrons
colliding and by the pair which does not collide) while only the  pair
 involved in the second and last collision produces a peak at $180^{\circ}$.
 Since the peaks at $90^{\circ}$ and $180^{\circ}$ 
 for the angular correlation of individual electron 
 pairs have roughly the same height (see \fig{fig5:li-angles-resolved})
 this should lead in the full angular correlation probability 
  to a peak at $90^{\circ}$ which is twice 
 as high as the one near  $180^{\circ}$. As one can see from
  \fig{fig4:li-angles-all} this is 
 indeed the case.

The T-shape is the consequence of the sequential process of two
collisions which dominates triple ionization since both main pathways
$s_{1}$ and $s_{2}$ contain two collisions.  Widely
unnoticed  the T-shape
configuration was already mentioned  in \cite{Grujic} for small
excess energy in a model-calculation of electron impact ionization of
Helium. The reason that this T-shape like escape has not received 
any attention since then lies probably in the fact that its 
connection to the fundamental organization of the triple escape in 
characteristic collision sequences of the electrons was not known 
until now. 

\subsection{Evolution of the angular correlation distribution with 
excess energy}
\begin{figure}
\centerline{\includegraphics[scale=0.8]{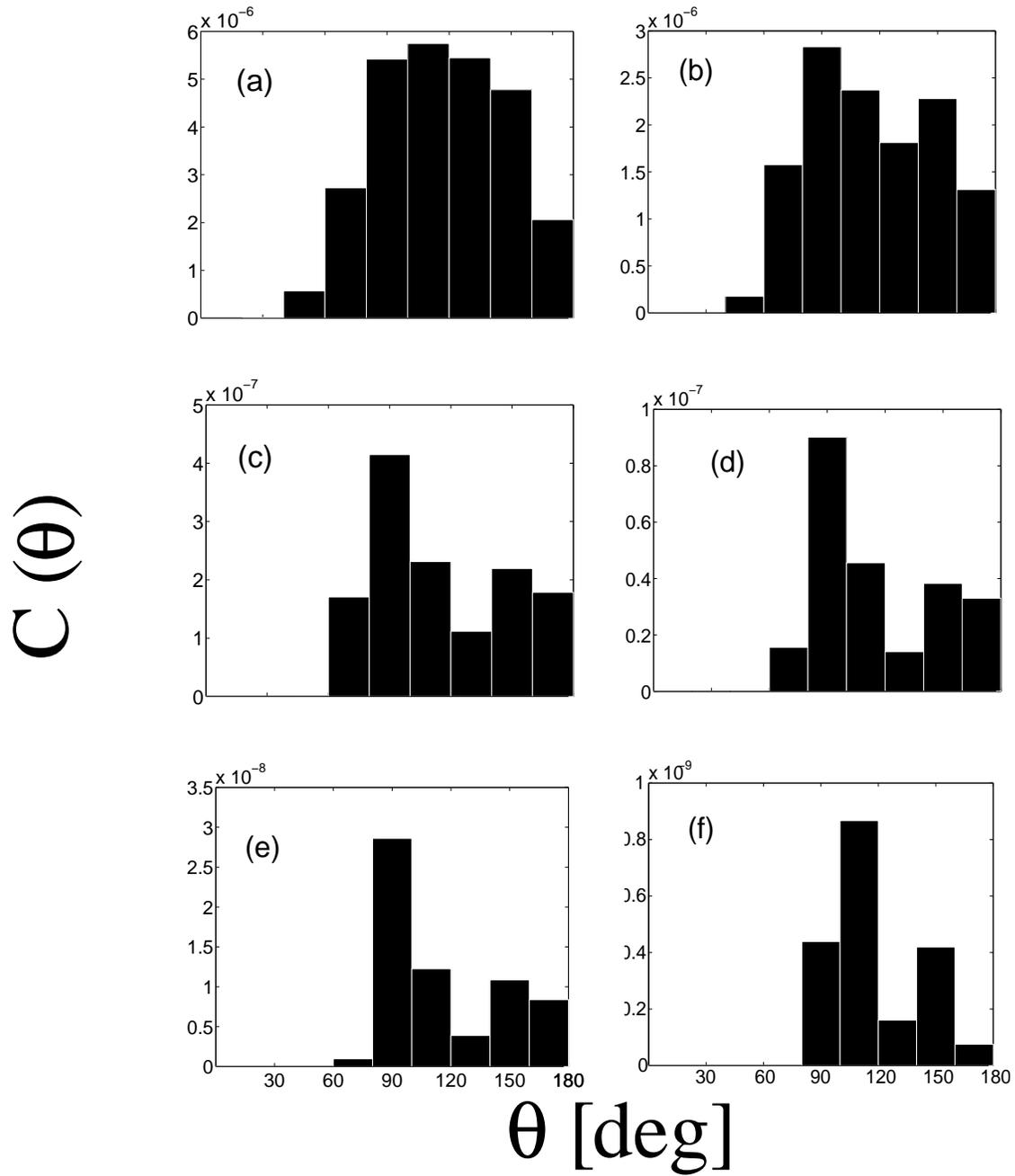}}
\caption{\label{fig2:transition}  Angular correlation probability $C(\theta)$
for excess 
energies of $E=$10 eV (a), 6.2 eV (b), 2 eV (c), 0.9 eV (d), 0.5 eV (e) 
and 0.1 eV (f). 
$C(\theta)$ has been binned over $20^{\circ}$ at
$10^{\circ},30^{\circ},\ldots,150^{\circ},170^{\circ}$ to reduce the 
statistical error. Its value for the total triple ionization probability
does not exceed 3\% in (a-e) and 10\% in (f).}
\end{figure}

As already indicated in figure \ref{fig4:li-angles-all} the T-shape 
is lost towards higher energies  where impulsive collisions dominate 
and the electrons are so fast that $\theta = 180^{\circ}$ is not 
reached.
On the other hand, approaching $E\to 0$ we expect the symmetric 
triangular escape according to Wannier.  
In our present calculation we see a clear tendency for the transition to the 
Wannier configuration only at the lowest excess energy (0.1 eV) we 
are able to calculate, see figure \ref{fig2:transition}f. This could mean that, 
at least concerning differential observables, the Wannier regime in 
triple ionization has an extremely short range in energy. It could, 
however, also mean that our approach, treating the photo electron 
very differently from the other electrons, favors asymmetric break up 
of the electrons and therefore a very low transition energy between 
symmetric and asymmetric electron configurations.

\section{Discussion and conclusions}
To summarize, we find  classically 
that in triple ionization of lithium by single
photon absorption the three electrons escape to the continuum through
a dominant T-shape configuration for excess energies that are in the
range of validity of the Wannier law ($\alpha=2.16$) \cite{emro06}.  This T-shape
configuration gives rise to a double hump structure in the
inter-electronic angular distribution of the escaping electrons at
$90^{\circ}$ and $180^{\circ}$.  We have explained this surprising
double hump structure in terms of a novel classification scheme
 that is built upon momentum transferring
electron-electron collisions.

 In the framework of many body perturbation theory (MPT) different
 ionization processes have been known for many years \cite{McGuire}.
 However, as the name already says, the perturbative character of this
 approach makes it applicable at high excess energies only with the
 electron--electron interaction treated to first or second order.  The
 classification scheme we have developed emerges from the full
 classical dynamics which includes the electron--electron interaction to
 all orders.  Surprisingly, our scheme shows that we can describe the
 triple photoionization process as a sequence of electron--electron
 collisions for energies close to threshold, where the Wannier theory
 becomes valid.

The sequence of collisions relevant for lithium involves only
three-body Helium-like subunits (nucleus and two electrons) at one
instant of time.  Since Coulomb systems interact via two-body forces
only, it may well be that this scheme holds for more than three
electrons in an atom.  At the same time this would imply that not a
two-electron atom but a three-electron atom is the fundamental system
whose understanding allows one to access multi--electron ionization
dynamics in the future.  Our scheme might also guide the way to a
quantum mechanical analysis along the lines of \cite{Bri90}.

Furthermore, our study may offer valuable insight into the triple
photoionization process in connection to double ionization by electron
impact.  For double ionization, the relationship between electron
impact ionization of He$^{+}$ and a quasiclassical formulation of the
double photoionization from the He ground state has already been
established \cite{Samson,scch+02}.  Both processes differ only
slightly, namely, in the energy scale set by the respective bound
electron.  We believe that our quasiclassical formulation of the
triple photoionization of the Li ground state is a process very
similar to double ionization of the excited states 1s2s $^{1,3}S$ of
Li$^{+}$ by electron impact.  The target state involved is the 1s2s
Li$^{+}$, since in the triple photoionization process the
photo-electron is a $1s$ electron knocking out the remaining two, 1s
and 2s, electrons.  Experimental/theoretical results on double
ionization of Li$^{+}$ by electron impact are needed to establish
whether or not the two three-electron escape processes are indeed
similar.  Such results would also elucidate how the spin
symmetry---not accounted for in our classical calculation---affects
the double ionization by electron impact from the excited states 1s2s
$^{1,3}S$ of Li$^{+}$.

 An experimental investigation of the double hump structure we predict
 for the inter-electronic angular distribution for small excess
 energies is very desirable. It could confirm the existence of the 
 T-shaped escape of the three electrons and would thereby support our 
 classification scheme of collisions as well as the validity and 
 limitations of a 
 classical approach to the four-body Coulomb problem.

 The authors gratefully acknowledge  Thomas Pattard for helpful discussions 
 and a critical reading of the manuscript.

\end{document}